\def\vector#1{\mbox{\boldmath ${#1}$}} %
\def\m#1{\mathrm{#1}} %\mathrm??????
\def\p{\partial} %\partial??????
\def\be#1{\begin{equation}#1\end{equation}} %{equation}
\def\beqn#1{\begin{eqnarray*}#1\end{eqnarray*}}
\def\beqnn#1{\begin{eqnarray}#1\end{eqnarray}}
\def\bra#1{\langle{#1}|} %?u???x?N?g??
\def\cket#1{|{#1}\rangle} %?P?b?g?x?N?g??
\documentclass{PoS}
\usepackage{comment}
\title{
\begin{picture}(0,0)(0,0)%
   \put(350,75){\makebox(0,0)[l]{\textnormal
{\normalsize KEK-CP-353}
}}%
\end{picture}%
Determination of charm quark mass from temporal moments of charmonium correlator with Mobius domain-wall fermion}

\ShortTitle{Determination of charm quark mass}

\author{\speaker{Katsumasa Nakayama}$^{a, b}$, Brendan Fahy$^b$, and Shoji Hashimoto$^{b,c}$ (JLQCD collaboration)\\
         $^a$ Department of Physics, Nagoya University, Nagoya, 464-8602, Japan\\
         $^b$ KEK Theory Center, High Energy Accelerator Research Organization (KEK), Tsukuba 305-0801, Japan\\
         $^c$ School of High Energy Accelerator Science, The Graduate University for Advanced Studies (Sokendai),Tsukuba 305-0801, Japan\\
        E-mail: \email{katumasa@post.kek.jp}}

\abstract{
We extract the charm quark mass and the strong coupling
constant
from the charmonium current correlators calculated with $n_f = 2 + 1$ Mobius
domain
wall fermions. 
We match our lattice calculation for the temporal moments of the correlator with perturbative result known up to four-loop order, and extract the charm quark mass with uncertainty less than 1\%.
Using the temporal moments,
we also confirm the correlators in the vector channel to be consistent with the experimental data for the R-ratio.
We used the ensembles generated by the JLQCD collaboration at lattice spacings $a = 0.080$ fm, 0.055 fm and 0.044 fm.}

\FullConference{34th annual International Symposium on Lattice Field Theory\\
		24-30 July 2016\\
		University of Southampton, UK}

\begin{document}

\section{Introduction}
Short-distance current correlators in QCD can be analyzed using
perturbation theory, while they can be directly calculated in lattice
QCD. By matching them, one may determine the parameters in the Standard
Model. The charm quark mass is a good example, {\it i.e.} it can be
extracted from the short-distance regime by means of the moment method
first proposed by the HPQCD-Karlsruhe collaboration \cite{Allison:2008xk}. The method
has also been used for the determination of the bottom quark mass by the
same group, and the precision has been improved \cite{Chakraborty:2014aca}. More recently,
we utilized the same method but with a different lattice formulation, 
to determine the charm quark mass \cite{Nakayama:2016atf}.

We use the lattice ensembles generated by the JLQCD collaboration with
the Mobius domain-wall fermion for 2+1 flavors of dynamical quarks. The
lattices are relatively fine, {\it i.e.} $a=0.080-0.055$ fm, 
which allow us to control the discretization effects.
In this talk, we mainly discuss a test of this method using 
experimental data, as well as the main sources of systematic uncertainty,
 while leaving the full description of this work in \cite{Nakayama:2016atf}.
The same set of lattice ensembles have also been used for the studies of
heavy-light decay constant \cite{BFahy} and semileptonic decay form factors \cite{Kaneko}.

For the vector channel, the current correlator
can be related to the $e^+e^-$ cross section, or the $R$ ratio, using
the optical theorem. By comparing lattice results with
phenomenological analysis obtained from experimental data, we may
validate the lattice calculation. We demonstrate that lattice data
are consistent with experiments after taking the continuum limit.

For the determination of the charm quark mass, we use the pseudo-scalar
channel, as it provides a more sensitive probe. Among other sources of
systematic uncertainty, including those of discretization effects and
finite volume effect and so on, it turned out that the perturbative
error is the dominant source. We attempt to conservatively estimate the
effect of perturbative error.
%%%%%%%%%%
%
\section{Moment of correlators}
We calculate the correlators of the pseudo-scalar current
$j_5 = i\bar{\psi_c}\gamma_5\psi_c$
and vector current
$j_k = \bar{\psi_c}\gamma_k\psi_c$
composed of charm quark field $\psi_c$:
\begin{eqnarray}
  \label{eq:G^PS}
  G^{PS}(t) & = & a^6 \sum_{\vector{x}} (am_c)^2 
  \langle 0| j_5 (\vector{x},t)j_5 (0,0) |0\rangle,
  \\
  \label{eq:G^V}
  G^{V}(t) & = & \frac{a^6}{3}\sum_{k=1}^3 \sum_{\vector{x}} 
  Z_V^2
  \langle 0| j_k (\vector{x},t)j_k (0,0) |0\rangle,
\end{eqnarray}
with a renormalization constant for the vector current $Z_V$.
We then construct the temporal moments
\beqn{
G_n & = & \sum_t \left(\frac{t}{a}\right)^n G(t),
}
\begin{comment}
\begin{eqnarray}
  \label{eq:momentPS}
  G_n^{PS} & = & \sum_t \left(\frac{t}{a}\right)^n G^{PS}(t),
  \\
  \label{eq:momentV}
  G_n^V & = & \sum_t \left(\frac{t}{a}\right)^n G^V(t),
\end{eqnarray}
\end{comment}
%
for each channel with an even number $n\geq 4$.
Since the charmonim correlators $G(t)$ are exponentially suppressed in the long-distance regime, the moments are sensitive to the region of $t\sim n/M$ depending on the charmonium ($\eta_c$ or $J/\psi$) mass $M$.

The moments are related to the vacuum polarization functions $\Pi ^V (q^2)$  and $\Pi ^{PS} (q^2)$ as
\begin{eqnarray}
  (q^\mu q^\nu-q^2g^{\mu\nu})\Pi^V(q^2)
 & = &
 i\int d^4x\, e^{iqx}
  \bra{0}j^\mu(x) j^\nu(0)\cket{0},
\\
  q^2 \Pi^{PS} (q^2)
  & = &
  i \int d^4x\, e^{iqx} 
  \langle 0| j_5(x)j_5(0)|0\rangle.
\end{eqnarray}
through the derivatives with respect to $q^2$:

\be{
a^{2k}G_{2k+2} ^V = \frac{12\pi^2 Q_f ^2}{k!}\left(\frac{\p}{\p q^2}\right)^k \left(\Pi^V(q^2)\right)|_{q^2 = 0}.
}
The vector channel can be related to the experimentally observed $e^+e^-$ cross section, {\it i.e.} the $R$-ratio $R(s)\equiv\sigma_{e^+e^-\to c\bar{c}}(s)/\sigma_{e^+e^-\to\mu^+\mu^-}(s)$ using the optical theorem:
\be{
\frac{12\pi^2 Q_f ^2}{k!}\left(\frac{\p}{\p q^2}\right)^k \left(\Pi^V(q^2)\right)|_{q^2 = Q_0 ^2}
\equiv
  \int_{s_0}^{\infty} ds \frac{1}{(s-Q_0 ^2)^{k+1}} R(s).
}
Here $Q_0$ is an arbitrally number and often set to $Q_0 = 0$.
We use this relation between the lattice calculation and experimental data for consistency check of the lattice calculation.
% Figure \ref{fig:exper} shows our lattice calculation produce consistent result with the experimental data.

The temporal moments for sufficiently small $n$ can be calculated perturbatively since they are defined in the short-distance regime.
The valuum polarization functions are represented with a dimensionless parameter
$z \equiv q^2/2m_c ^2(\mu)$ as
\be{
\Pi(q^2) = \frac{3}{16\pi^2}\sum_{k=-1} ^{\infty}C_kz^k,
}
and the coefficients $C_k$ are perturbatively calculated up to $O(\alpha_s ^3)$ in the $\overline{\m{MS}}$ scheme \cite{Maier:2007yn,Maier:2009fz,Kiyo:2009gb}, 
%
\begin{comment}
\begin{eqnarray}
  C_k &=& C_k^{(0)} + \frac{\alpha_s(\mu)}{\pi} 
  \left( C_k^{(10)}+ C_k^{(11)}l_{m}\right) 
  \nonumber\\
  & & + \left( \frac{\alpha_s(\mu)}{\pi}\right) ^2 
  \left( C_k^{(20)} + C_k^{(21)}l_{m} + C_k^{(22)}l_{m}^2 \right) 
  \nonumber\\
  & & + \left( \frac{\alpha_s(\mu)}{\pi}\right)^3 
  \left( C_k^{(30)} + C_k^{(31)}l_{m} + C_k^{(32)}l_{m}^2 +
    C_k^{(33)}l_{m} ^3 \right) + ...,
\end{eqnarray}
\end{comment}
and written in terms of $l_m \equiv \m{log}(m^2 _c(\mu)/\mu^2)$ and $\alpha_s (\mu)$. Since we use this perturbative expansion to extract the charm quark mass and the strong coupling constant, the uncertainty of  $O(\alpha_s ^4)$ remains.

Practically, we redefine the moments to reduce the uncertainty from the scale setting as well as that from the leading discretization effect:
\beqnn{
  \label{eq:reduced_moment}
  R_n ^{PS}
  &= 
      \displaystyle
      \frac{am_{\eta_c}}{2a\tilde{m}_c}
      \left(\frac{G_n ^{PS}}{{G_n ^{PS} }^{(0)}}\right)^{1/(n-4)}
      & 
       \mbox{for}\;\; n\geq 6.\\
  \label{eq:reduced_moment_V}
  R_n^V 
  &= 
      \displaystyle
      \frac{am_{J/\psi}}{2a\tilde{m}_c}
      \left(\frac{G_n^V}{G_n^{V(0)}}\right)^{1/(n-2)} 
      &
       \mbox{for}\;\; n\geq 4.
}
with the pole mass of the domain-wall fermion $\tilde{m}_c$ and the tree level moment $G_n ^{(0)}$. We will use these reduced moments to test the consistency with experimental data, and to determine the quark mass and strong coupling constant.
\section{Consistency with experimental data}

Before discussing the extraction of the charm quark mass, we try to validate the lattice calculation using the vector channel together with the experimental data available for the $R$-ratio.

Our lattice ensembles are generated with $2 + 1$ flavors of Moebius domain-wall fermion at lattice spacings $a$ = 0.080, 0.055, and 0.044 fm. The spacial size $L/a$ is 32, 48, and 64 respectively, and the temporal size $T/a$ is twice as long as $L/a$. Three defferent values of bare charm quark mass are taken to calculate charmonium correletors, and they are interpolated to the physical point such that the mass of spin-averaged 1S states are reproduced. The details of the ensembles are in \cite{BFahy}. The
renormalization constant $Z_V$ is determined non-perturbatively from the light hadron correlators as 0.955(9), 0.964(6), and 0.970(5) for $\beta$ = 4.17, 4.35, and 4.47, respectively \cite{Tomii:2016xiv}.

We extrapolate the data for $R_n ^V$ to the continuum limit using an ansatz
\begin{equation}
  R_n^V = R_n^V(0) 
  \left( 1 + c_1(am_c) ^2 \right) \times 
  \left( 1 + f_1\frac{m_u + m_d + m_s}{m_c} \right),
\label{fittingfunc}
\end{equation}
with three free parameters $R_n^V(0)$, $c_1$, and $f_1$. 
Higher order terms of $a$ and $m_l$ are confirmed to be
insignificant from the data.
We consider five different sources of uncertainty.
They are statistical error, finite volume effect, discretization error, uncertainty of the renormalization constant, and dynamical charm quark correction.
Since we use $2+1$ flavor ensembles in the lattice calculation, the dynamical charm quark effect is included using perturbation theory \cite{Maier:2007yn,Maier:2009fz,Kiyo:2009gb}.

\begin{figure}[tbp]
\begin{center}
  \includegraphics[width=6.5cm, angle=-90]{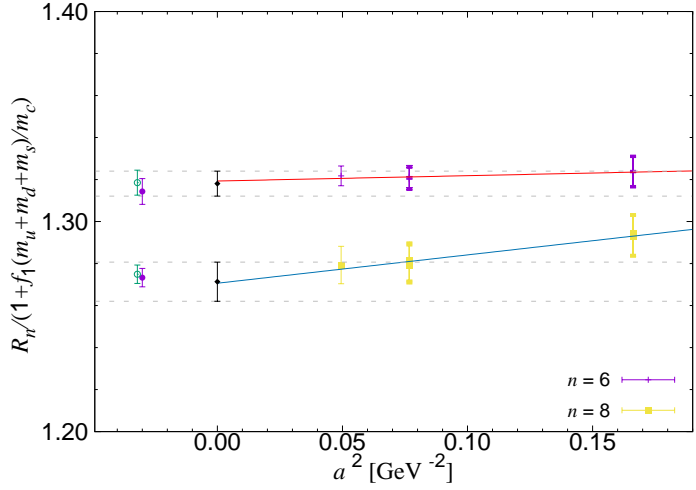}
  \caption{
    Reduced moments for the vector
    current $R_n^V$ ($n$ = 6 (pluses) and 8 (squares))
    and their continuum extrapolation.
    Data are plotted after correcting for the finite light quark mass
    effects by multiplying $1/(1+f_1(m_u+m_d+m_s)/m_c)$
    and for the missing charm quark
    loop effect 
    $r_n^V(n_f=4)/r_n^V(n_f=3)$.
    Phenomenological estimates of the corresponding quantities 
    are plotted on the left: 
    Dehnadi {\it et al.} \cite{Dehnadi:2011gc} (filled circle), 
    Kuhn {\it et al.} \cite{Kuhn:2007vp} (open circle). 
%    Kuhn {\it et al.} \cite{Kuhn:2001dm} (filled square), 
  %  and Hoang {\it et al.} \cite{Hoang:2004xm} (open square).
  }
\label{fig:vectorex}
\end{center}
\end{figure}

The result is shown in Figure \ref{fig:vectorex}.
The ``experimental data'' are taken from \cite{Dehnadi:2011gc,Kuhn:2007vp}, which are obtained by integrating the experimentally observed $R(s)$ with appropriate weight functions.
The lattice results show only mild $a$ dependence for $n$ = 6 and 8, and their continuum limit
is consistent with the corresponding ``experimental data''.
The dominant source of error is the renormalization constant, and the combined error is about 1\%,
which is about the same in size with the phenomenological estimate.
This agreement gives confidence about the validity of our lattice calculation.
%This consistency check means our spectrum in typical scale $\sim M / n$ from lattice calculation is reliable and its precision is comparable with experiment. After this confirmation, we can provide the moment of any channel not only vector channel with same precision, and we use pseudo-scalar channel to determine charm quark mass.

\section{Charm quark mass extraction}
\begin{figure}[tbp]
\begin{center}
  \includegraphics[width=6.5cm,angle=-90]{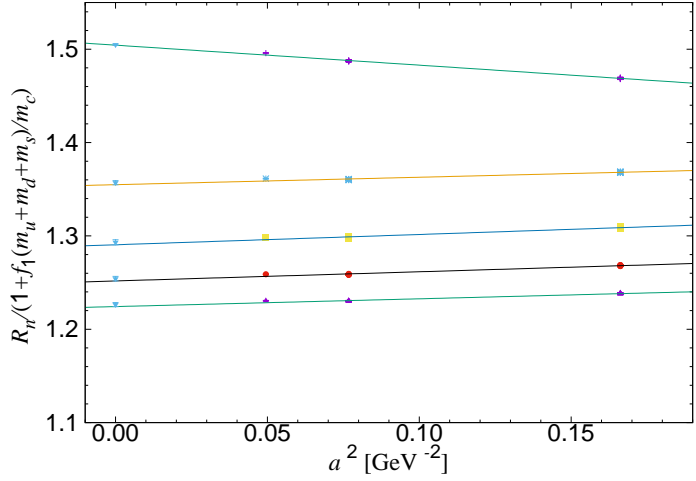}
  \caption{
    Continuum extrapolation of $R_n(a)$.
    Data points correspond to $R_6$, $R_8$, $R_{10}$, $R_{12}$, and $R_{14}$
    from top to bottom.
    We plot the mean of the extrapolation with and without coarsest lattice as the extrapolated values,
   and estimate it deviation as the $O(a^4)$ error.
  }
  \label{fig:R_n extrap}
\end{center}
\end{figure}

We use the reduced moment $R_n$ of the pseudo-scalar channel to determine the charm quark mass.
The continuum extrapolation of $R_n$ is shown in Figure \ref{fig:R_n extrap} with statistical error.
We assume the extrapolation form to be the same as that of $R_n ^V$ (\ref{fittingfunc}) with free parameters $R_n (0)$, $c_1$, and $f_1$, and use the perturbative factor $r_n(n_f=4)/r_n(n_f=3)$ to correct for the charm sea quark contribution. Our extrapolated lattice data are sufficiently precise since they have small lattice spacing $a$ dependence.
%The error from perturbative expantion have to be discussed more precisely, so here is not quoted.

Now we consider the systematic error from the perturbative expansion for the reduced moments $r_n$, which are known up to $O(\alpha _s ^3)$ \cite{Maier:2007yn,Maier:2009fz,Kiyo:2009gb}, and the leading uncertainty is at the order of $\alpha _s ^4$.
Such error from unknown higher order terms can be estimated by residual $\mu$ dependence of the perturbative result, since the physical quantity should be independent of the renormalization scale $\mu$.
We choose the range $\mu = 2-4$ GeV to estimate this source of error.
Below the lower limit the perturbative result rapidly varies, which suggests that the perturbative expansion is no longer reliable.
In the moment method, the combination $r_n(\alpha_s(\mu),m_c(\mu))/m_c(\mu)$ has to be $\mu$ independent.

We generalize this procedure for the scales to define $\alpha_s (\mu)$ and of $m_c(\mu)$ separately. Namely, we use the perturbative expansion written in terms of $\alpha_s(\mu_\alpha)$ and $m_c(\mu_m)$ with $\mu_\alpha \neq \mu_m$ \cite{Dehnadi:2015fra}. We estimate the truncation error using the range $\mu_\alpha \in \mu_m \pm 1$ GeV with 2 GeV $\leq \m{min}\{\mu_\alpha,\mu_m\}$ and $\m{max}\{\mu_\alpha,\mu_m\}\leq$ 4 GeV.
By allowing the possibility of $\mu _\alpha \neq \mu _m$, the estimated error becomes twice as large.
We adopt this choice to be conservative.
%This extention produce twice error as the truncation uncertainty.
%
%
%
%
%
%
%

The contribution from the gluon condensate, which appears in the operator product expansion of $r_n$, is another source of error. It can be written as
\be{
g_{2k} ^{GG} = \frac{\langle(\alpha_s/\pi)G_{\mu\nu} ^2\rangle}{2m_c(\mu)}\left(a_l + \frac{\alpha_s}{\pi}c_l\right),
}
where the coefficients $a_l$ and $c_l$ are known up to $O(\alpha_s ^2)$ \cite{Broadhurst:1994qj}.
The gluon condensate $\langle(\alpha_s/\pi)G_{\mu\nu} ^2\rangle$
is not well determined phonomenologically, {\it e.g.} $\langle(\alpha_s/\pi)G_{\mu\nu} ^2\rangle = 0.006 \pm 0.0012\m{\ GeV}^4$ from a $\tau$ dacay analysis \cite{Geshkenbein:2001mn}.
In our analysis, we treat  $\langle(\alpha_s/\pi)G_{\mu\nu} ^2\rangle$ as a free parameter and determine from the lattice data together with $m_c(\mu)$ and $\alpha_s(\mu)$.

In the deffinition of the moments, there appears a meson mass $m_{\eta_c}$, which is an input parameter.
Because our lattice calculation does not contain the electromagnetic and disconnected diagram effects, we need to modify the mass of $\eta_c$ to take account of their effects. The electromagnetic effects is expected to reduce the meson mass by 2.6(1.3) MeV \cite{Davies:2009tsa}, and the disconected contribution also reduces the mass by 2.4(8) MeV according to a lattice study \cite{Follana:2006rc}. We therefore use the modified $m_{\eta_c}$ as an input, $m_{\eta_c} ^{\m{exp}}= 2983.6(0.7) + 2.4(0.8)_\m{Disc.} + 2.6(1.3)_\m{EM}$ MeV.
\begin{table}
\begin{center}
  \begin{tabular}{c|c|cccccccc}
    \hline
    & & pert &$t_0 ^{1/2}$& stat & $O(a^4)$ & vol & $m_{\eta_c}^{\m{exp}}$ & disc & EM\\
    \hline
%    $R_6$, $R_8$, $R_{10}$ & 
%    1.0032(98) & (82) &(51) & (5) & (16) & (4) & (3) & (4) & (6)\\
%    $R_6$, $R_6/R_8$, $R_{10}$ & 
%    1.0031(194) & (176) &(78) & (6) & (18) & (5) & (4) & (4) & (7)\\
    $m_c(\mu)$ [GeV]&1.0033(96)\ \ \ \ & (77)\  \ &(49) & (4) & (30) & (4) & (3) & (4) & (6)\\
    \hline 
  \end{tabular}

  \vspace{0.5pt}  
  \begin{tabular}{c|c|cccccccc}
    \hline
%    $R_6$, $R_8$, $R_{10}$ & 
%    0.2530(256) & (213) &(134) & (12) & (38) & (10) & (9) & (10) & (16)\\
%    $R_6$, $R_6/R_8$, $R_{10}$ &
%    0.2528(127)	& (120)	&(33) & (2) & (25) & (1) & (0) & (0) & (1)\\
  $\alpha_s(\mu)$\ \ \ \ \ \ \ \ \ \ \ \ &  0.2528(127)\ \ & (120) &(32) & (2) &\ (26)	&\ \ (1) &\ (0) &\ \ (0) &\ \  (1)\\
    \hline
  \end{tabular}

  \vspace{0.5pt}
  \begin{tabular}{c|c|cccccccc}
    \hline
%    $R_6$, $R_8$, $R_{10}$ &
  %  $-$0.0005(99) & (85) &(45) & (4) & (23) & (4) & (3) & (4) & (6) \\
%    $R_6$, $R_6/R_8$, $R_{10}$ &
%    $-$0.0006(144) & (133) &(49) & (4) & (23)	& (4) & (3) & (3) & (5)\\
    $\frac{<(\alpha / \pi)G^2>}{m^4}$\ \ \ \ \ \ & $-$0.0006(78)\ \ & (68) &(29) & (3) &\ (22) &\ \ (3) &\ (2) &\ \  (3) &\ \  (5)\\
    \hline 
  \end{tabular}
  \caption{
    Numerical results for $m_c(\mu)$ (top panel), $\alpha_s(\mu)$ (mid
    panel) and $\frac{<(\alpha_s/\pi)G^2>}{m^4}$ (bottom panel) 
    at $\mu$ = 3~GeV.
    The results are listed for choices of three input
    quantities out of $R_8$, $R_{10}$ and $R_6/R_8$.
    In addition to the central values with combined errors, the
    breakdown of the error is presented.
    They are the estimated errors from the truncation of perturbative
    expansion, the input value of $t_0 ^{1/2}$, statistical, discretization error of $O(a^4)$ (or
    $O(\alpha_sa^2)$),
    finite volume, experimental data for $m_{\eta_c}^{\m{exp}}$,
    disconnected contribution, electromagnetic effect, in the order
    given. 
    The total error is estimated by adding the individual errors in
    quadrature. 
  }
  \label{nogluin2}
  \end{center}
\end{table}

\begin{figure}[tbp]
  \centering
  \includegraphics[width=8.9 cm, angle=0]{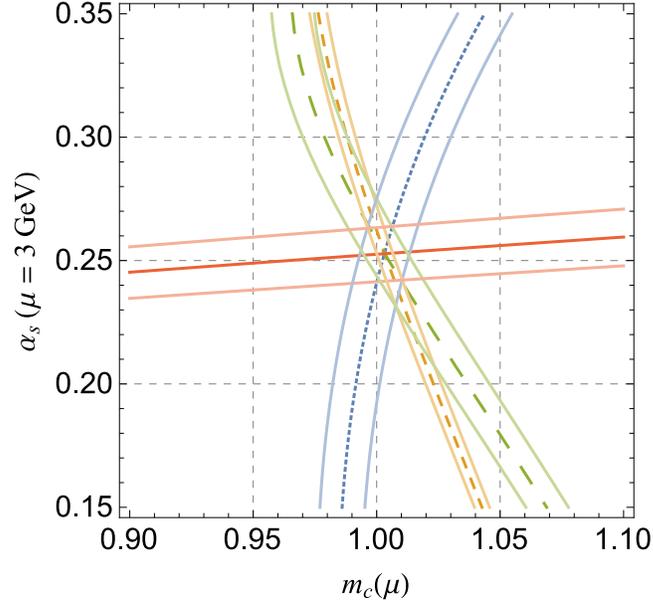}
  \caption{
    Constraints on $m_c(\mu)$ and $\alpha_s(\mu)$
    from the moments 
    $R_6$ (dotted curve), $R_8$ (dashed curve), 
    $R_{10}$ (long dashed curve), and $R_6/R_8$ (solid curve).
    For each curve, the band represents the error due to the
    truncation of perturbative expansion.
  }
  \label{fig:noingraph}
\end{figure}

We include all of these error estimates.
Namely, statistical error, discretization effect of $O(a^4)$, finite volume, experimental value of $m_{\eta_c} ^\m{exp}$, disconnected and electromagnetic effect. Table \ref{nogluin2} lists the result of charm quark mass $m_c(\mu)$ and strong coupling $\alpha_s(\mu)$ as well as the  gluon condensate $\langle(\alpha_s/\pi)G_{\mu\nu} ^2\rangle$ in the $\overline{\m{MS}}$ scheme at $\mu = 3$ GeV.
Figure \ref{fig:noingraph} shows the constraints on $m_c$ and $\alpha_s$ from the moments and their ratio. Since each moment puts different constraints on these parameters, charm quark mass $m_c(\mu)$ and coupling constant $\alpha_s (\mu)$ can be determined.
Roughly speaking, the individual moment is more sensitive to $m_c (\mu)$ while the ratio $R_6/R_8$
has a sensitivity to $\alpha_s(\mu)$.

In the final result, the dominant source of the error comes from the truncation of perturbative expansion for all quantities.
The next largest is the discretization effect of $O(a^4)$ as well as the uncertainty of lattice scale determined with the Wilson flow $t_0 ^{1/2}$.
It means that in order to achive more precise determination with this method,
we need yet another order of perturbative expansion.
\vspace{17pt}
The lattice QCD simulation has been performed on Blue Gene/Q supercomputer at the High Energy Accelerator Research Organization (KEK) under the Large Scale Simulation Program (Nos. 13/14-4, 14/15-10, 15/16-09). This work is supported in part by the Grant-in-Aid of the Japanese Ministry of Education (No. 25800147, 26247043, 26400259).
%
%
%
%
%
% 
%\begin{thebibliography}{99}
%\bibitem{...}
%....
%\end{thebibliography}

%%% Local Variables:
%%% mode: latex
%%% TeX-master: t
%%% End:

\end{document}